\documentclass[12pt]{article}
\usepackage{amsmath,amsbsy,amssymb,graphics}
\usepackage{graphicx,epsfig}
\def\beq{\begin{equation}}
\def\eeq{\end{equation}}
\def\be{\begin{equation}}
\def\ee{\end{equation}}
\def\bea{\begin{eqnarray}}
\def\eea{\end{eqnarray}}
\def\nnb{\nonumber}

\newcommand{\gsim}{\lower.7ex\hbox{$\;\stackrel{\textstyle>}{\sim}\;$}}
\newcommand{\lsim}{\lower.7ex\hbox{$\;\stackrel{\textstyle<}{\sim}\;$}}

\begin{document}

\begin{flushright}
\baselineskip=12pt
MIFP-08-05 \\
\end{flushright}

\begin{center}
 \vspace{0.2cm}
 {\Large \bf  Generalization of Friedberg-Lee Symmetry }

\vspace{0.6cm} {\large \bf Chao-Shang Huang$^c$, Tianjun Li$^{c,d}$,
 Wei Liao$^{a,b}$\footnote{Corresponding author, email: liaow@ecust.edu.cn },
 Shou-Hua Zhu$^e$}

\vspace{0.3cm} { $^a$ Institute of Modern Physics,\\
 East China University of Science and Technology, \\
 P.O. Box 532, 130 Meilong Road, Shanghai 200237, P.R. China\\

 \vskip 0.3cm

 $^b$ Center for High Energy Physics,\\
 Peking University, Beijing 100871, P. R. China\\

 \vskip 0.3cm

 $^c$ Institute of Theoretical Physics, Chinese Academy of
Sciences, \\
       P. O. Box 2735, Beijing 100080, P. R. China \\
       \vskip 0.3cm

$^d$ George P. and Cynthia W. Mitchell Institute for
Fundamental Physics,  \\
Texas A$\&$M University, College Station, TX
77843, USA \\

\vskip 0.3cm

 $^e$ Institute of Theoretical Physics, School of Physics, \\
  Peking University, Beijing 100871, P. R. China}

\end{center}
\begin{abstract}
 \vskip 0.2cm
We study the possible origin of Friedberg-Lee symmetry. First, we
propose the generalized Friedberg-Lee symmetry in the potential by
including the scalar fields in the field transformations, which can
be broken down to the FL symmetry spontaneously. We show that the
generalized Friedberg-Lee symmetry allows a typical form of Yukawa
couplings, and the realistic neutrino masses and  mixings can be
generated via see-saw mechanism. If the right-handed neutrinos
transform non-trivially under the generalized Friedberg-Lee
symmetry, we can have the testable TeV scale see-saw mechanism.
Second, we present two models with the $SO(3)\times U(1)$ global
flavour symmetry in the lepton sector. After the flavour symmetry
breaking, we can obtain the charged lepton masses, and explain the
neutrino masses and mixings via see-saw mechanism. Interestingly,
the complete neutrino mass matrices are similar to those of the
above models with generalized Friedberg-Lee symmetry. So the
Friedberg-Lee symmetry is the residual symmetry in the neutrino mass
matrix after the $SO(3)\times U(1)$ flavour symmetry breaking.

\end{abstract}


\section{Introduction}\label{sec1}

 Recent developments in neutrino physics ~\cite{skatm,sno1,sno2,sk,chooz,kamland}
 have stimulated many interesting new ideas~\cite{BSV,ms,lam,FL1}.
 One beautiful approach
 towards understanding neutrino masses and mixings was presented
 by Friedberg and Lee~\cite{FL1,FL2,FL3}.
 They showed that there may be a hidden symmetry
 in the neutrino mass matrix with tri-bimaximal mixings, {\it i.e.},
the invariance under
 the translation in the space of Grassmann number
 \bea
 \nu_{e,\mu,\tau} \to \nu_{e,\mu,\tau}+\theta. \label{FL1}
 \eea
 The symmetry was later used to explain the quark masses and mixings~\cite{FL2}.
Instead of a universal translation for all fermions,
 they introduced different coefficients in translation of different flavors
 of quarks
 \bea
 q_i \to q_i +\xi_i \theta. \label{FL2}
 \eea
 And this symmetry implies that one family of the Standard Model (SM)
fermions is massless. Explicit symmetry breaking terms are introduced to
reproduce the masses for the light SM fermions. Researches along
 this approach have been performed by several groups~\cite{Xing,Jarlskog}.

 On the other hand, it is generally acknowledged that the see-saw
mechanism~\cite{Min,yana,grs,glas,MohSen} is a powerful method to
understand the tiny masses of the active neutrinos.
See-saw mechanism needs a symmetry
to guarantee the masslessness of the neutrinos at leading order. The masses of
light neutrinos are generated after symmetry breaking. In this respect it
is natural to ask what kind symmetry can implement the see-saw mechanism
in such a way that the Friedberg-Lee (FL) symmetry is the residual symmetry
hidden in the neutrino mass matrix.

 In this article, we first generalize the FL symmetry in a simple way
by including the scalar fields in the left-handed neutrino field transformations.
 The generalized Friedberg-Lee (gFL) symmetry naturally incorporates
the FL symmetry. And the  FL symmetry of Eq. (\ref{FL1}) or Eq. (\ref{FL2})
is obtained after the larger gFL symmetry breaking.
 The masslessness of three light
 neutrinos is a direct consequence of the gFL symmetry. After the gFL
 symmetry is broken down to FL symmetry, the light neutrinos get masses via
see-saw mechanism, and
 their masses and mixings are intimated related to the residual FL symmetry.
We show that the observed neutrino masses and mixings can be reproduced
via see-saw mechanism. Also, if the transformations of the right-handed neutrinos
under the gFL symetry is similar to those of the left-handed neutrinos,
 the testable TeV scale see-saw mechanism can be realized.
Moreover, we briefly discuss how to embed the models with gFL symmetry
into the extensions of the SM.
Second, we propose two models with the $SO(3)\times U(1)$
global flavour symmetry in the lepton sector. After the flavour symmetry breaking,
the charged lepton masses
can be obtained, and the neutrino masses  and mixings can be
generated via see-saw mechanism .
Interestingly, the complete neutrino mass matrices for the left-handed
and right-handed neutrinos are similar
to those of the above models with gFL symetry. So the FL symmetry is the residual
symmetry  in the neutrino mass matrix
after the $SO(3)\times U(1)$ flavour symmetry breaking.

 The content of this article is organized as follows. In Section
 \ref{sec2} we propose the gFL symmetry and study
the models with gFL symmetry. And in Section 3, we consider
the models with $SO(3)\times U(1)$ flavour symmetry in the lepton
sector. Our conclusions and discussions are in Section \ref{sec6}.

\section{Generalized Friedberg-Lee Symmetry} \label{sec2}

We consider two models with the generalization of  FL symmetry. One
model has usual see-saw mechanism where only  the left-handed
neutrinos transform non-trivially under the gFL symmetry, and the
other model has the testable TeV scale see-saw mechanism in which
both the left-handed and right-handed neutrinos transform
non-trivially under the gFL symmetry.

\subsection{Usual See-Saw Mechanism}
\label{sec2.1}

 We consider three families of the left-handed neutrinos
 $\nu_{Li}$, right-handed neutrinos $\nu^c_{Ri}$
 and three SM singlet scalar fields $\phi_i$ where $i=1,2,3$.
 We introduce the generalized Friedberg-Lee symmetry by
 including scalar fields in the field transformations of $\nu_{Li}$.
 We introduce the following gFL symmetry transformation
 \bea
 && \nu_{Li} \to \nu_{Li}+ ~\phi_i ~\theta~,~~~
  \nu^c_{Ri} \to \nu^c_{Ri}~,~~\phi_i \to \phi_i~,~\,
 \label{trans1}
 \eea
 where $\theta$ is a Grassmann number\footnote{This gFL symmetry
 is introduced for neutrinos after electroweak breaking. One may consider
 that $\theta$ carries an isospin number. Extension to doublet is discussed
 in section \ref{sec2.4} }.
 We require that the neutrino mass terms and Yukawa
 terms be invariant under this symmetry transformation.

 The FL symmetry is obtained after the
 gFL symmetry breaks spontaneously. This can be achieved by
 assuming the potential of $\phi_i$ triggers
 the spontaneous symmetry breaking. We assume
 $\phi_i$ have a potential as follows
 \bea
 -\Delta {\cal L}= \xi~ (\sum^3_{i=1} |\phi_i|^2-v^2)^2 ~,~\,  \label{vacuum1}
 \eea
 where $\xi >0$. Then, $\phi_i$ get the  vacuum expectation values (VEVs)
at the minimum of the potential
 \bea
 <\phi_i > = v_i \label{vacuum2} ~,~\,
 \eea
 where $v^2=\sum_{i=1}^{3} |v_i|^2$. 
 The induced transformation is as follows
 \bea
 && \nu_{Li} \to \nu_{Li}+ ~v_i ~\theta, ~~~
 \nu^c_{Ri} \to \nu^c_{Ri} \label{trans2}~.~\,
 \eea
Because the coefficients $v_i$ in above equation are
 space-time independent, we obtain the FL symmetry as a
residual symmetry.

  The mass term and Yukawa terms invariant under the gFL
 transformation are
\begin{eqnarray}
 -\Delta {\cal L} &=& \frac{1}{2} (m_0)_{ij} ~\nu^{cT}_{R i} ~i \sigma_2 \nu^c_{R j}
 + \lambda_{ijk} ~\nu^{cT}_{R i} ~i \sigma_2 ~\nu_{L j} ~\phi_k
 +\frac{1}{2} \eta_{ijk} ~\nu^{cT}_{R i} ~i \sigma_2 ~\nu^c_{R j} ~\phi_k
 \nonumber\\
 && +\frac{1}{2} \eta^{\prime}_{ijk} ~\nu^{cT}_{R i} ~i \sigma_2
~\nu^c_{R j} ~\phi^{\dagger}_k + h.c. ~,~\,
 \label{poten}
 \end{eqnarray}
 where $\lambda_{ijk}$, $\eta_{ijk}$
 and $\eta^{\prime}_{ijk}$ are Yukawa couplings, and
 $\nu^T$ means the transpose of $\nu$. Also, we have to impose
 \bea
 \lambda_{ijk}=-\lambda_{ikj}~,~ \label{symm1}
 \eea
 \bea
 (m_0)_{ij}=(m_0)_{ji}~, ~~\eta_{ijk}=\eta_{jik}~,~~
\eta^{\prime}_{ijk}=\eta^{\prime}_{jik}~.~\,  \label{symm2}
 \eea
 The first, the third and the fourth terms in Eq. (\ref{poten})
are obviously invariant under
 the gFL transformation in Eq. (\ref{trans1}).
 Eq. (\ref{symm1}) is required to make the second term invariant under
 the gFL transformation.
 Using Eq. (\ref{symm1}) the second term transforms to
 \bea
 && \lambda_{ijk} ~\nu^{cT}_{R i} ~i \sigma_2 ~\nu_{L j} \phi_k
 + \lambda_{ijk} ~\nu^{cT}_{R i} ~i \sigma_2 ~\theta ~ \phi_j~ \phi_k \nnb \\
 &&= \lambda_{ijk} ~\nu^{cT}_{R i} ~i \sigma_2 ~\nu_{L j}
 \phi_k ~,~\,
 \eea
 So it is invariant under the gFL symmetry.
 However, the other terms, {\it e.g.}, $\nu^T_{L i} i \sigma_2 ~ \nu_{L j}$ and
 $\nu^T_{L i} ~i \sigma_2 ~\nu_{L j} \phi_k$, etc, are not invariant
 under the gFL transformation and are killed by the gFL symmetry.

 We see that the mass term $\nu^T_{L i} i \sigma_2 ~ \nu_{Lj}$
 is killed by the gFL symmetry defined in Eq. (\ref{trans1}).
 If gFL symmetry is not broken to the FL symmetry neutrinos $\nu_{Li}$
 won't be able to get masses.
 In this sense the masslessness of three $\nu_{Li}$
 is a direct consequence of the gFL symmetry. Neutrinos $\nu_{Li}$
 get see-saw type masses after $\phi_i$ get vevs and
 gFL symmetry in Eq. (\ref{trans1}) is broken to the residual FL symmetry
 in Eq. (\ref{trans2}). The generation of the see-saw masses for $\nu_{Li}$
 is shown in the following.

 After the gFL symmetry is broken down to the FL symmetry, we obtain
the following neutrino mass terms
 \bea
 -\Delta {\cal L}= \frac{1}{2}(m_R)_{ij} ~\nu^{cT}_{R i} ~i \sigma_2 \nu^c_{R j}
 + \Lambda_{ij} ~\nu^{cT}_{R i} ~i \sigma_2 ~\nu_{L j}  + h.c. ~,~\,
 \label{mass1}
 \eea
 where
 \bea
 (m_R)_{ij}=(m_0)_{ij}+\sum_k \eta_{ijk} v_k
+ \sum_k \eta^{\prime}_{ijk} v^*_k~.~\, \label{mass2}
 \eea
\bea
\Lambda_{ij}= \sum_k \lambda_{ijk} ~v_k ~.~\,
\label{mass3b}
 \eea

 It is obvious that Eq. (\ref{mass1}) is invariant under the residual
 FL symmetry transformation in Eq. (\ref{trans2}). And
 we can write the neutrino mass matrix in the basis
 $(\nu_L,\nu^c_R)^T$ as follows
 \bea
 {\cal M}=\begin{pmatrix} 0_{3\times 3}~~, & ~~\Lambda^T \cr \Lambda~~, &
 ~~~  m_R           \end{pmatrix} ~,~\,  \label{mass3}
 \eea
 where $\Lambda$ and $m_R$ are $3\times 3$ matrices, and
their matrix elements are $\Lambda_{ij}$ and $(m_R)_{ij}$,
respectively.

 Assuming the mass scale of $\Lambda$ is much lower than that of $m_R$
 we get the see-saw mass matrix for the light neutrinos
 \bea
 m_{\nu} =- \Lambda^T ~(m^{-1}_R) ~\Lambda ~.~\,  \label{mass4}
 \eea

 Thus, using the gFL symmetry we have implemented see-saw mechanism.
 It is clear that the gFL symmetry protects the masslessness of
 neutrinos $\nu_L$. Right-handed neutrinos $\nu^c_R$ are allowed
 to have masses and are heavy. Only one typical form of the neutrino
 Dirac Yukawa couplings
 is allowed by the gFL symmetry. This type of the Yukawa couplings introduces
 the mixings of $\nu_L$ and $\nu^c_R$. After the gFL symmetry
 is spontaneously broken down to the FL symmetry we get a see-saw type mass
 matrix for $(\nu_L,\nu^c_R)^T$ and the see-saw mass matrix for
 the light neutrinos, which are shown in Eqs. (\ref{mass3}) and
 (\ref{mass4}), respectively.

 \subsection{Neutrino Masses and Mixings}\label{sec2.2}

 In this subsection we give examples which can reproduce the realistic neutrino masses
 and mixings. For simplicity we assume $v_i$ are real and
 $\Lambda$ and $m_R$ are real matrices.
 For illustration we will try to obtain the following
 tri-bimaximal neutrino mixing matrix~\cite{hps,hs,Xing2}
 \bea
 U=\begin{pmatrix} \sqrt{\frac{2}{3}} & \sqrt{\frac{1}{3}} & 0 \cr
  -\sqrt{\frac{1}{6}} & \sqrt{\frac{1}{3}} & \sqrt{\frac{1}{2}} \cr
  \sqrt{\frac{1}{6}} & - \sqrt{\frac{1}{3}} & \sqrt{\frac{1}{2}}
  \end{pmatrix} ~,~\, \label{Numixing}
 \eea
 which has $\theta_{13}=0$, $\theta_{23}=\pi/4$ and $\tan^2\theta_{12}=0.5$.
 More realistic textures can be done by following the discussions
 in this subsection.

 A direct consequence of the residual FL symmetry in Eq. (\ref{trans2})
 is that one light neutrino is massless. This can be seen by noting
 that under the transformation
 $\nu_{Li} \to \nu_{Li} + v_i \theta$ the see-saw mass term
 of the light neutrinos is transformed to (after rearrangement)
 \bea
 && \frac{1}{2}(m_\nu)_{ij} \nu_{Li}^T i \sigma_2 \nu_{Lj} +h.c.\to \nnb \\
 && \to \frac{1}{2} (m_\nu)_{ij} [\nu_{Li}^T i \sigma_2 \nu_{Lj}
 +  2 v_j \nu_{Li}^T i \sigma_2 \theta
 +  v_i v_j \theta^T i \sigma_2 \theta ]+h.c. \label{masslessb}
 \eea
 The invariance under the FL symmetry
 transformation says that the second term in the bracket of the r.h.s. of Eq.
 (\ref{masslessb}) gives zero. Hence we obtain
 \bea
  m_{\nu} ~\begin{pmatrix} v_1 \cr v_2 \cr v_3 \end{pmatrix} =0.
 \label{massless}
 \eea
 So neutrinos $\nu_{Li}$ have one eigenstate with zero mass.
 The eigenvector is $(v_1,v_2,v_3)^T$.
 \footnote{A difference between the gFL symmetry and the FL symmetry is that the
 coefficients $\phi_i$ in the transformation law of the gFL symmetry
 (hence the coefficients in the eigenvector) can take arbitrary values,
 instead of fixed constants $v_i$ in the FL transformation.
 So gFL symmetry makes three neutrinos massless and the FL
 symmetry only guarantees one neutrino massless.}
 Eq. (\ref{massless}) can also be obtained by using
 Eqs. (\ref{mass3b}), (\ref{mass4}) and (\ref{symm1}) directly.

We shall present two examples. The first example has inverted hierarchy.
For simplicity, we assume that
 $m_R$ is a unit matrix, {\it i.e.}, $m_R= m_s {\bf 1}$. And
 we choose
 \bea
 (v_1,v_2,v_2)^T=\frac{v}{\sqrt{2}}(0,1,1)^T~.~\,  \label{choice1}
 \eea

 Using Eq. (\ref{choice1}) we get
 \bea
 m_{\nu}=-\frac{v^2}{2 m_s} \begin{pmatrix}
  F_2, & F_{\lambda}, & -F_{\lambda} \cr
  F_{\lambda}, & \lambda_2, & - \lambda_2 \cr
 -F_\lambda, & -\lambda_2, & \lambda_2
  \end{pmatrix}~,~\, \label{Numass1}
 \eea
 where
 \bea
 F_2=\sum_i (\lambda_{i12}+\lambda_{i13})^2~,~~
 F_\lambda=\sum_i \lambda_{i23} (\lambda_{i12}+\lambda_{i13})~,~~
 \lambda_2= \sum_i \lambda_{i23}^2~.~\,
  \label{Numass1b}
 \eea
 We find that $m_{\nu}$ is diagonalized by $U$
 \bea
 U^T ~m_{\nu} ~U=-\frac{v^2}{2 m_s} \textrm{diag}
 \{F_2-F_\lambda,F_2+2 F_\lambda,0\}~,~\,
 \label{Numass2}
 \eea
 provided that the following condition is satisfied
 \bea
 F_2+F_\lambda =2 \lambda_2~.~\, \label{cond1}
 \eea
 And we get
 \bea
 \Delta m^2_{21}=3 F_\lambda (2 F_2 +F_\lambda) \frac{v^4}{4 m_s^2}~, ~~\Delta
 m^2_{31}=-(F_2-F_\lambda)^2 \frac{v^4}{4 m_s^2}~.~\,
 \eea
 The realistic neutrino mass square differences can be obtained since
 we have enough independent parameters to fit two $\Delta m^2$.

 The second example has normal hierarchy, we take $\Lambda$
 anti-symmetric and $m_R$ diagonal
 \bea
 m_R=\textrm{diag}\{m_{r1},m_{r2},m_{r3}\}~.~\,
 \eea
 We choose
 \bea
 (v_1,v_2,v_3)^T=\frac{v}{\sqrt{6}} ( 2,-1,1)^T~.~\, \label{choice2}
 \eea
 Using $\lambda_{ijk}=\lambda \epsilon_{ijk}$ we get
 \bea
 \Lambda = \frac{v}{\sqrt{6}}
 \begin{pmatrix} 0 & \lambda & \lambda \cr
  -\lambda & 0 & 2 \lambda \cr
  -\lambda & -2 \lambda & 0
 \end{pmatrix}~.~\,
 \eea
 And we find
 \bea
 m_{\nu} =-\frac{\lambda^2 v^2}{6}
 \begin{pmatrix}
 \frac{1}{m_{r2}}+\frac{1}{m_{r3}}, & \frac{2}{m_{r3}}, & -\frac{2}{m_{r2}} \cr
 \frac{2}{m_{r3}}, & \frac{1}{m_{r1}}+\frac{4}{m_{r3}}, & \frac{1}{m_{r1}} \cr
 -\frac{2}{m_{r2}}, & \frac{1}{m_{r1}}, & \frac{1}{m_{r1}}+\frac{4}{m_{r2}}
 \end{pmatrix} \label{Numass3} ~.~\,
 \eea
 If the condition
 \bea
 m_{r2}=m_{r3} \label{cond2}
 \eea
 is satisfied, we find
 \bea
 U^T ~m_\nu~U =- \frac{\lambda^2 v^2}{3} \textrm{diag}\{0,
 \frac{3}{m_{r2}}, \frac{1}{m_{r1}}+\frac{2}{m_{r2}} \}~.~\, \label{Numass4}
 \eea
 Hence we get
 \bea
 \Delta m^2_{21}=\frac{\lambda^4 v^4}{m^2_{r2}},~~
 \Delta m^2_{31}= \frac{\lambda^4 v^4}{9} (\frac{1}{m_{r1}}
 +\frac{2}{m_{r2}})^2.
 \eea
 Using the hierarchy in neutrino mass $\Delta m^2_{31} \approx 25
 \Delta m^2_{21}$ we find
 \bea
 m_{r2} \approx 13 m_{r1}.
 \eea

 \subsection{Testable TeV Scale See-Saw Mechanism}\label{sec2.3}

 In recent years there have been some interests in the TeV
 scale see-saw mechanism~\cite{TeVSS1,TeVSS2}. The mechanism suggests that the
 mixings of the left-handed and right-handed neutrinos are
 independent of the hierarchy in the Dirac type and Majorana
 type masses. This makes the see-saw mechanism testable at the future
 colliders or in rare decay processes.
 In this subsection we show that we can also realize the testable TeV
 scale see-saw mechanism via  the generalized Friedberg-Lee symmetry.

 Instead of Eq. (\ref{trans1}) we introduce the following gFL
 symmetry transformation under which the right-handed neutrinos
transform non-trivially as well
 \bea
 && \nu_{Li} \to \nu_{Li}+ \frac{1}{\sqrt{1+|\alpha_i|^2}}~\phi_i ~\theta~,~\, \label{trans3} \\
 && \nu^c_{Ri} \to \nu^c_{Ri}+ \frac{\alpha_i}{\sqrt{1+|\alpha_i|^2}} ~\phi_i ~\theta~,~\,
 \label{trans4} \\
 && \phi_i \to \phi_i ~,~ \label{trans4b}
 \eea
  where $\alpha_i~(i=1,2,3)$ are complex numbers.

  We introduce neutrinos $\nu_{\bot}$ and
 $\nu_{\top}$ in an orthogonal basis
 \bea
  \nu_{\bot i} =\frac{1}{\sqrt{1+|\alpha_i|^2}} (\nu_{Li}+\alpha_i^*~
  \nu^c_{Ri}), \label{def1} \\
  \nu_{\top i} =\frac{1}{\sqrt{1+|\alpha_i|^2}} (-\alpha_i~
  \nu_{Li}+\nu^c_{Ri})~.~\,  \label{def2}
 \eea
 It is easy to see that under Eqs. (\ref{trans3}) and (\ref{trans4})
 we have
 \bea
 \nu_{\bot i} \to \nu_{\bot i}+\phi_i~\theta,~~
 \nu_{\top i} \to \nu_{\top i} ~.~\,  \label{trans5}
 \eea
Thus, in the
 new basis the Eq. (\ref{trans1}) is reproduced.  And then the discussions on the
 see-saw mechanism and the neutrino masses and mixings are
  similar to those in the subsections \ref{sec2.1} and \ref{sec2.2}.
 The only difference
 with the previous case is that the mixings between the left-handed and right-handed
 neutrinos are no longer suppressed by the mass hierarchy in the see-saw type
 mass matrix in Eq. (\ref{mass3}). Denoting the neutrino mass eigenstates
 as $(\nu,\nu_H)^T$ we can find that
 \bea
 \begin{pmatrix} \nu_L \cr \nu^c_R \end{pmatrix}
 \approx \begin{pmatrix}  {\cal A}_0, & - {\cal A}^\dagger_1 \cr
 {\cal A}_1, & {\cal A}_0 \end{pmatrix}
 \begin{pmatrix}
  U, &  0 \cr 0, & U_H
 \end{pmatrix}
 \begin{pmatrix} \nu \cr \nu_H \end{pmatrix} \\ \nnb
 = \begin{pmatrix}
 {\cal A}_0 ~U , & -{\cal A}^\dagger_1 ~U_H \cr
 {\cal A}_1 ~U , & {\cal A}_0 U_H \end{pmatrix}
 \begin{pmatrix}
 \nu  \cr \nu_H
 \end{pmatrix}, \label{mixing}
 \eea
 where $U$ is the mixing matrix of the light neutrinos $\nu$,
 $U_H$ is the mixing matrix of heavy neutrinos $\nu_H$, and
 \bea
 && {\cal A}_0 =\textrm{diag} \{{1 \over \sqrt{1+|\alpha_1|^2} },
 {1 \over \sqrt{1+|\alpha_2|^2} }, {1 \over \sqrt{1+|\alpha_3|^2} } \}~,~~ \label{mixinga} \\
 && {\cal A}_1 =\textrm{diag} \{ {\alpha_1 \over \sqrt{1+|\alpha_1|^2} },
 {\alpha_2 \over \sqrt{1+|\alpha_2|^2} }, {\alpha_3 \over \sqrt{1+|\alpha_3|^2} }\} ~.~\,
\label{mixingb}
 \eea
 We find  that the mixings of the left-handed and right-handed neutrinos are
 determined by $\alpha_i$ which is independent of the mass hierarchy
 between the Dirac type and Majorana type masses. $A_1$ determine
 the strength of unitarity violation of the mixings of light neutrinos~\cite{xing3}.
 This kind scenario may be possibly tested at the future colliders~\cite{hz}
 and neutrino oscillation experiments~\cite{go}.

\subsection{Embedding into the Extensions of the SM} \label{sec2.4}

We can embed the above models into the extensions of the SM. Let us
denote the SM lepton doublets as $L_i$, and the SM Higgs field as
$H$. Also, we introduce three SM singlet scalar fields $\phi_i$. By
the way, the following discussions can be easily generated to the
supersymmetric Standard Models by changing \bea H \to H_u~,~~~
{\widetilde H} \to  H_d~,~ \, \eea where ${\widetilde H} =i\sigma_2
H^*$, and $H_u$ and $H_d$ are
one pair of the Higgs doublets in the supersymmetric Standard Models.\\

(A) For the usual see-saw mechanism, we introduce
the following  gFL symmetry
\bea
 && L_i \to L_i+ ~{\phi_i} ~\chi~,~~
  \nu^c_{Ri} \to \nu^c_{Ri}~,~~\phi_i \to \phi_i~,~~H \to H~,~\,
\eea
 where $\chi$ is an $SU(2)_L$ doublet and has two components
 of Grassmann constant. And the relevant neutrino Lagrangian is
\begin{eqnarray}
-\Delta {\cal L} &=&   \frac{1}{2} (m_0)_{ij} ~\nu^{cT}_{R i} ~i
\sigma_2 \nu^c_{R j}+
 \lambda_{ijk} \overline{\nu}_{Ri} L_j  {{\phi_k}\over {M_*}} H
 +\frac{1}{2} \eta_{ijk} ~\nu^{cT}_{R i} ~i \sigma_2 ~\nu^c_{R j} ~\phi_k
\nonumber\\&&
+\frac{1}{2} \eta^{\prime}_{ijk} ~\nu^{cT}_{R i} ~i \sigma_2 ~\nu^c_{R j}
~\phi^{\dagger}_k + H.C. ~,~\,
\label{EESM-Lagrangian-A}
\end{eqnarray}
 where $\lambda_{ijk}=-\lambda_{ikj}$, and $(m_0)_{ij}$, $\eta_{ijk}$
 and $\eta^{\prime}_{ijk}$ are symmetric for $i$ and $j$,
 and $M_*$ is the cutoff scale of the gFL symmetry.
 Because the Lagrangian in Eq. (\ref{EESM-Lagrangian-A})
 is similar to that in Eq. (\ref{poten}),
 we can embed the model with the usual see-saw mechanism into
 the extension of the SM.

 As a remark, the most naive approach is that we introduce
 three Higgs doublets $H_i$, and define the following  gFL symmetry
\bea
 && L_i \to L_i+ ~{\widetilde H_i} ~\theta~,~~
  \nu^c_{Ri} \to \nu^c_{Ri}~,~~H_i \to H_i~,~\,
\eea
 where ${\widetilde H}_i =i\sigma_2 H_i^*$.
 However, the neutrino Dirac Yukawa couplings
 $\overline{\nu}_{Ri} L_j H_k$ are not invariant under the
 above gFL symmetry. And then
 we can not explain the neutrino masses and mixings
 via see-saw mechanism. In short, this approach does not
 work. \\

 (B) For the testable TeV scale see-saw mechanism, we
 have to embed the three right-handed neutrinos into
 three fermionic doublets $L'_i$. To cancel the anomaly,
 we introduce three fermionic doublets ${\widetilde L}'_i$
 which are the Hermitian conjugate of $L'_i$. And  we introduce
 the  gFL symmetry transformation as follows
\bea
 && L_i \to L_i+ \frac{1}{\sqrt{1+|\alpha_i|^2}}
~{\phi_i}~\chi~,~\,  \\
 && L'_i \to L'_i+ \frac{\alpha_i}{\sqrt{1+|\alpha_i|^2}}
 ~{\phi_i} ~\chi~,~\, \\
 && \phi_i \to \phi_i~,~~H \to H ~,~~~ {\widetilde L}'_i \to  {\widetilde L}'_i ~.~ \,
\eea
  And we define \bea
  L_{\bot i} =\frac{1}{\sqrt{1+|\alpha_i|^2}} (L_i+\alpha_i^*~
  L'_i), \label{def1b} \\
  L_{\top i} =\frac{1}{\sqrt{1+|\alpha_i|^2}} (-\alpha_i~
  L_i+L'_i)~.~\,  \label{def2b}
 \eea
 It is easy to see that under the above gFL symmetry, we have
 \bea
 L_{\bot i} \to L_{\bot i}+ {\phi_i} ~\chi,~~
 L_{\top i} \to L_{\top i} ~.~\,  \label{trans5b}
 \eea
And then we obtain
 the major relevant neutrino Lagrangian
\begin{eqnarray}
-\Delta {\cal L} &=& {1 \over {M_I}} \left(\lambda^{\nu}_{ijkl}
L_{\bot i} L_{\bot j} {{\phi_k}\over {M_*}} {{\phi_l}\over {M_*}}
H^2 +y^{\nu}_{ijk} L_{\top i} L_{\bot j}  {{\phi_k}\over {M_*}} H^2
+\lambda_{ij} {\widetilde L}'_i {\widetilde L}'_j {\widetilde H}
{\widetilde H} \right) \nonumber\\&& + M_{ij} L_{\top i} {\widetilde
L}'_j + y^L_{ijk} L_{\top i} {\widetilde L}'_j \phi_l + H.C.~,~\,
\label{Double-SS}
\end{eqnarray}
 where $M_I$ is an intermediate scale and the Yukawa couplings
 $\lambda^{\nu}_{ijkl}$ satisfy $\lambda^{\nu}_{ijkl} = -\lambda^{\nu}_{kjil}
 =-\lambda^{\nu}_{ilkj}$ or $\lambda^{\nu}_{ijkl} = -\lambda^{\nu}_{ljki}
 =-\lambda^{\nu}_{ikjl}$, and the Yukawa coupling $y^{\nu}_{ijk}$ is anti-symmetric
 for  $j$ and $k$.
 Interestingly, the neutrino mass matrix proposed
 by Friedberg and Lee can be generated by the first term in Eq.
 (\ref{Double-SS}). Even if this term is zero, {\it i.e.},
 $\lambda^{\nu}_{ijkl}=0$, the observed neutrino masses and mixings
 can be generated by the double see-saw mechanim~\cite{extended, Kang:2004ix}. Here,
 we emphasize that we neglect the other
 high-dimensional operators that are not important in the discussions
 of the neutrino masses and mixings.

 In addition, the first three terms in Eq. (\ref{Double-SS}) are
 non-renormalizable and can be obtained by the see-saw mechanism.
  For example, if we introduce three SM
singlet fermions $N_i$, the first three terms can be obtained due to
the following Lagrangian via the see-saw mechanism
 \begin{eqnarray}
 -\Delta {\cal L} &=& {1\over 2} (M_N)_{ij}
 \overline{N}^c_{i} N_{j} +
 \lambda_{li} \overline{N}_{l} L_{\top i}  H +
 \eta_{ljk} \overline{N}_{l} L_{\bot j}  {{\phi_k}\over {M_*}}
 H +
 \lambda_{li} N_{l}  {\widetilde L}'_i {\widetilde H} + H.C.~,~\,
 \end{eqnarray}
where $(M_N)_{ij}$ is symmetric, and $\eta_{ljk}=-\eta_{lkj}$.
 $M_I$ is around the mass scales of $N_i$.

\section{$SO(3)\times U(1)$ Flavour Symmetry in the Lepton Sector}

To explain the SM fermion masses and mixings, we usually use the
Froggatt-Nielsen mechanism~\cite{Froggatt:1978nt} by introducing the
global flavour symmetry. Thus, the FL symmetry could also be a
residual symmetry after the flavour symmetry breaking. In this
section, we consider the $SO(3)\times U(1)$ flavour symmtry in  the
lepton sector.

Let us explain the convention in details. We denote the SM Higgs doublet as $H$,
 the left-handed lepton doublets as $L_i$, and right-handed
charged leptons as $E_i$. To break the $SO(3)\times U(1)$ flavour
symmetry we also introduce three Higgs doublets $H_i$, and nine SM
singlet scalar field $\Phi$, $\Phi_i$ and $\Phi_{ij}$. We assume
that the $L_i$, $E_i$,  $H_i$ and $\Phi_i$  form the fundamental
representation of $SO(3)$, and $\Phi_{ij}$ form the symmetric
representation of $SO(3)$. We shall present two concrete models in
the following subsections: In the Model I, $\nu_{Ri}$ are singlets
under $SO(3)$, while in Model II, $\nu_{Ri}$ form the fundamental
representation of $SO(3)$ and we do not need the $\Phi_i$ fields.

\subsection{FL symmetry with see-saw mechanism}
 Before we study the $SO(3)\times U(1)$ flavour Symmetry, let us consider the
FL model with see-saw mechanism.  We consider the FL symmetry as
follows
 \bea
  L_i \to L_i+ ~ \xi_i ~\chi~,~~\nu_{Ri} \to \nu_{Ri}~,~~H \to H ~,~ \,
\label{Intro-1}
\eea
 where we obtain the original FL symmetry by
choosing $\xi_1=\xi_2=\xi_3$. And the neutrino Lagrangian, which is
invariant under above FL symmetry, is
\begin{eqnarray}
-\Delta {\cal L} &=&  {1\over 2} (m'_0)_{ij} \overline{\nu}^c_{Ri}
\nu_{Rj} + y_{ijk} \overline{\nu}_{Ri} (\xi_k L_j-\xi_j L_k) H ~.~
\, \label{Intro-AA} \eea Following the usual
procedure~\cite{Min,yana,grs,glas,MohSen}, we realize the see-saw
mechanism with FL symmetry in the light neutrino mass matrix.
Therefore, in order to generalize the FL symmetry, we need to
construct the models that can reproduce the above Lagrangian in Eq.
(\ref{Intro-AA}) after the generalized symmetry breaking. As an
example to explain the main idea, we introduce three SM Higgs
doublets and consider $\xi_i H$ as $H_i$. Then the above neutrino
Lagrangian becomes

\begin{eqnarray}
-\Delta {\cal L} &=&  {1\over 2} (m'_0)_{ij} \overline{\nu}^c_{Ri}
\nu_{Rj} + {1\over 2} (m'_0)_{ij} \overline{\nu}^c_{Ri} \nu_{Rj}  +
y_{ijk} \overline{\nu}_{Ri} (L_j H_k- H_j L_k)  ~.~ \,
\label{Intro-LAG} \eea
 Therefore, we can obtain the neutrino mass matrix with FL
symmetry if the neutrino Dirac Yukawa couplings $y_{ijk} $ are
anti-symmetric for the lepton doublet indices $j$ and Higgs field
indices $k$,  {\it i.e.}, $y_{ijk}=-y_{ikj}$.

\subsection{Model I}

We assume that under
 the $U(1)$ symmetry, $\nu_{Ri}$  has charge 0,
 $L_i$ has charge $1$,   $E_i$ has charge $-1/2$, $H$ has charge $1/2$,
$H_i$ has charge 2, $\Phi_i$ has charge $-3$, and $\Phi$ and
$\Phi_{ij}$ have charges $-1$. The $SO(3)\times U(1)$ invariant
Lagrangian is
\begin{eqnarray}
-\Delta {\cal L} &=&  {1\over 2} (m'_0)_{ij} \overline{\nu}^c_{Ri}
\nu_{Rj} + {1\over {M_{\rm Pl}}} \left( y^{\nu}_{ijkl}
\overline{\nu}_{Ri} L_j H_k \Phi_l + \lambda^{E} \overline{E}_i L_i
{\widetilde H} \Phi \right.\nonumber\\&&\left. +y^{E}_{ij}
\overline{E}_i L_j {\widetilde H} \Phi_{ij} \right) + H.C.~,~ \,
\label{SO(3)-Lagrangian}
\end{eqnarray}
where the Yukawa couplings $y^{\nu}_{ijkl}$ are anti-symmetric for
their indices $j$, $k$, and $l$ due to the $SO(3)$ invariance. For
simplicity, we assume that the SM Higgs field $H$ has VEV close to
174 GeV, while  the Higgs fields  $H_i$ have small VEVs, for
example, a few GeVs. In addition, we assume that $\Phi$, $\Phi_i$
and $\Phi_{ij}$ have VEVs around the grand unification
 scale $2.4\times 10^{16}$ or higher so that the dimension-5 operators
can generate the masses for the charged leptons and neutrinos.
And it is not difficult to show that
we do have enough degrees of freedom to
explain the charged lepton masses, and the neutrino masses and mixings.

After the $SO(3)\times U(1)$ flavour symmetry breaking,   we obtain that
the  neutrino mass matrix for the left-handed and right-handed
neutrinos from the Lagrangian in Eq. (\ref{SO(3)-Lagrangian}) is
the same as that from the Lagrangian  in Eq. (\ref{poten})
by choosing the following relations
\begin{eqnarray}
(m_0)_{ij}+\eta_{ijk} \langle \phi_k \rangle +\eta^{\prime}_{ijk}
\langle \phi^*_k \rangle = (m'_0)_{ij} ~,~ \lambda_{ijk} \langle
\phi_i \rangle = {1\over {M_{\rm Pl}}}
 y^{\nu}_{ijkl} \langle H_k \rangle \langle  \Phi_l \rangle~.~\,
\end{eqnarray}
Similar to the discussions in the subsection 2.2, we can explain the
realistic neutrino masses and mxings. Interestingly, the
$SO(3)\times U(1)$ flavour symmetry is broken down to the FL
symmetry. In other words, the FL symmetry is the residual symmetry
in the neutrino mass matrix from the flavour symmetry breaking.

Moreover, the FL symmetry can be broken only by the dimension-7 or
higher operators. And the dimension-7 operators that break the FL
symmetry are
\begin{eqnarray}
-\Delta {\cal L} &=& {1\over {M^3_{\rm Pl}}} \overline{\nu}_{Ri} L_j
H_k (\Phi^3 \delta_{jk}+ \Phi^2 \Phi_{jk} + \Phi_{jl} \Phi_{lm}
\Phi_{mk} + \Phi_{jk} \Phi_{lm} \Phi_{lm}) + H.C.~,~\,
\end{eqnarray}
where for simplicity we neglect the Yukawa couplings.
Thus, the FL symmetry is a very good approximate symmetry in
the neutrino mass matrix.

By the way, the VEVs of $\Phi$, $\Phi_i$ and $\Phi_{ij}$ break the
$U(1)$ symmetry down to the $Z_2$ symmetry. Under this $Z_2$
symmetry, $E_i$ and $H$ are odd while the other fields are even. And
then, this $Z_2$ symmetry forbids the Dirac Yukawa couplings between
$H$ and neutrinos. Otherwise, the discussions will become very
complicated because the VEV of $H$ is much larger than those of
$H_i$ while the VEVs of $\Phi$, $\Phi_i$, and $\Phi_{ij}$ are close
to the Planck scale. Also, this $U(1)$ symmetry will not affect the
quark Yukawa couplings if we assign the $U(1)$ charges $1/2$ and
$-1/2$ to the right-handed up-type and down-type quarks,
respectively.

\subsection{Model II}

We assume that under
 the $U(1)$ symmetry, $\nu_{Ri}$  has charge $-1$,
 $L_i$ has charge $1$,   $E_i$ has charge $-3/2$, $H$ has charge $1/2$,
$H_i$ has charge $-2$,  and $\Phi$ and $\Phi_{ij}$ have charges
$-2$. The $SO(3)\times U(1)$ invariant Lagrangian is
\begin{eqnarray}
-\Delta {\cal L} &=&  {1\over 2} \lambda^N \overline{\nu}^c_{Ri}
\nu_{Ri} \Phi^{\dagger} + {1\over 2} y^N_{ij} \overline{\nu}^c_{Ri}
\nu_{Rj} \Phi_{ij}^{\dagger} + y^{\nu}_{ijk} \overline{\nu}_{Ri} L_j
H_k \nonumber\\&& + {1\over {M_{\rm Pl}}} \left( \lambda^{E}
\overline{E}_i L_i  {\widetilde H} \Phi +y^{E}_{ij} \overline{E}_i
L_j {\widetilde H} \Phi_{ij} \right)+ h.c.~,~ \,
\label{SO(3)-Lagrangian-2}
\end{eqnarray}
where  the Yukawa coupling $y^{\nu}_{ijk}$ is anti-symmetric for
their indices $i$, $j$ and $k$. Similar to the above subsection, we
have enough degrees of freedom to explain the charged lepton masses.

After the $SO(3)\times U(1)$ flavour symmetry breaking,   we obtain
that the  neutrino mass matrix for the left-handed and right-handed
neutrinos from the Lagrangian in Eq. (\ref{SO(3)-Lagrangian-2}) is a
special case of that from the Lagrangian  in Eq. (\ref{poten}) by
choosing the following relations
\begin{eqnarray}
(m_0)_{ij}+\eta_{ijk} \langle \phi_k \rangle +\eta^{\prime}_{ijk}
\langle \phi^*_k \rangle = \lambda^N \langle \Phi^{\dagger} \rangle
\delta_{ij} +y_{ij}^N  \langle \Phi^{\dagger}_{ij} \rangle~,~
\lambda_{ijk} \langle  \phi_i \rangle =
 y^{\nu}_{ijk} \langle H_k \rangle ~.~\,
\end{eqnarray}
The point is that the Yukawa coupling $y^{\nu}_{ijk}$ is
anti-symmetric for  $i$, $j$ and $k$ while $\lambda_{ijk}$ is only
anti-symmetric for  $j$ and $k$. Similar to the second example in
the subsection 2.2, we can explain the observed neutrino masses and
mixings. And  the FL symmetry is the residual symmetry from the
$SO(3)\times U(1) $ flavour symmetry breaking as well. Unlike the
Model I, it is very difficult to break the FL symmetry via the
higher dimensional operators, so the FL symmetry may be a symmetry
in the neutrino mass matrix.

\section{Conclusions and Discussions} \label{sec6}

In summary, we study the possible origin of the FL symmetry.
First, we  generalize the FL symmetry to the gFL symmetry
by including the scalar fields in the field transformations.
And the  FL symmetry is the residual symmetry after
 the larger gFL symmetry breaking.
 A direct consequence of the gFL symmetry is the masslessness of
three light neutrinos, which obtain masses via see-saw mechanism
after the gFL symmetry breaking. We also show that the observed
neutrino masses and mixings can be generated. Also, if the
transformations of the right-handed neutrinos under the gFL symmetry
are similar to those of the left-handed neutrinos, we can have the
testable TeV scale see-saw mechanism. Moreover, the models with gFL
symmetry can be embedded into the extensions of the SM. Second, we
propose two models with the $SO(3)\times U(1)$ global flavour
symmetry in the lepton sector. After the flavour symmetry breaking,
we can obtain the charged lepton masses, and explain the neutrino
masses and mixings via see-saw mechanism. In particular, the
complete neutrino mass matrices are similar to  those of the above
models with gFL symetry. So, the $SO(3)\times U(1)$ flavour symmetry
is broken down to the FL symmetry which is the residual symmetry
 in the neutrino mass matrix.

\section*{Acknowledgment}

 The research is supported in part by National
 Science Foundation of China (NSFC) under grant number
 10745003, 10775001, 10635030, 10435040, and by the
Cambridge-Mitchell Collaboration in Theoretical Cosmology (TL).

\end{document}